# Discovering One Molecule Out of a Million: Inverse Design of molecular hole transporting semiconductors tailored for perovskite solar cells

Jianchang Wu[1,2*†], Luca Torresi[3,4†], ManMan Hu[5†], Patrick Reiser[3,4†], Jiyun Zhang[1,2], Juan S. Rocha-Ortiz[1,2], Luyao Wang[6*], Zhiqiang Xie[2], Kaicheng Zhang[2], Byung-wook Park[5], Anastasia Barabash[1,2], Yicheng Zhao[1,2,7], Junsheng Luo[2,7], Yunuo Wang[2], Larry Lüer[1,2], Lin-Long Deng[6], Jens A. Hauch[1,2], Sang Il Seok[5], Pascal Friederich[3,4*], Christoph J. Brabec[1,2,8*]

[1]Forschungszentrum Jülich GmbH, Helmholtz-Institute Erlangen−Nürnberg (HI-ERN), 91058 Erlangen, Germany
[2]Faculty of Engineering, Department of Material Science, Materials for Electronics and Energy Technology (i-MEET), Friedrich-Alexander-Universität Erlangen−Nürnberg (FAU), 91058 Erlangen, Germany
[3]Institute of Nanotechnology, Karlsruhe Institute of Technology (KIT), 76131 Karlsruhe, Germany
[4]Institute of Theoretical Informatics, Karlsruhe Institute of Technology (KIT), 76131 Karlsruhe, Germany
[5]Department of Energy Engineering, School of Energy and Chemical Engineering, Ulsan National Institute of Science and Technology (UNIST), Ulsan 44919, Korea
[6]State Key Lab for Physical Chemistry of Solid Surfaces, Department of Chemistry, College of Chemistry and Chemical Engineering, Pen-Tung Sah Institute of Micro-Nano Science and Technology, Xiamen University, Xiamen, 361005 China
[7]National Key Laboratory of Electronic Films and Integrated Devices, School of Integrated Circuit Science and Engineering, University of Electronic Science and Technology of China, 611731 Chengdu, P. R. China
[8]Zernike Institute for Advanced Materials, University of Groningen, Groningen 9747 AG, The Netherlands.

***Corresponding authors.** E-mail: jianchang.wu@fau.de (J.W.); wangluyaoyz@gmail.com (L.W.); pascal.friederich@kit.edu (P.F.); christoph.brabec@fau.de (C.J.B.)
†These authors contributed equally to this work.

**Abstract**
The inverse design of tailored organic molecules for specific optoelectronic devices of high complexity holds an enormous potential, but has not yet been realized[1,2]. The complexity and literally infinite diversity of conjugated molecular structures present both, an unprecedented opportunity for technological breakthroughs as well as an unseen optimization challenge. Current models rely on big data[3,4], which do not exist for specialized research films. However, a hybrid computational and high throughput experimental screening workflow allowed us to train predictive models with as little as 149 molecules. We demonstrate a unique closed-loop workflow combining high throughput synthesis and Bayesian optimization that discovers new hole transporting materials with tailored properties for solar cell applications. A series of high-performance molecules were identified from minimal suggestions, achieving up to 26.23% (certified 25.88%) power conversion efficiency in perovskite solar cells. Our work paves the way for rapid, informed discovery in vast molecular libraries, revolutionizing material selection for complex devices. We believe that our approach can be generalized to other emerging fields and indeed accelerate the development of optoelectronic semiconductor devices in general.

**Main**

The evolution of organic semiconductors has historically been driven by empiricism. The design of hole-transporting materials (HTMs) for perovskite solar cells (PSCs), where experimentalists attempt to qualitatively recognize patterns in HTM structures to improve device performance, is no exception [5-7]. Nevertheless, this approach is challenged by various factors, including a lack of mechanistic understanding for new HTMs and the inherent limitations of human cognition in recognizing patterns in high-dimensional datasets. The rise of big-data techniques has the potential to revolutionize materials discovery. Machine learning is the study and construction of computer algorithms capable of extracting valuable insights and making predictions from vast sets of data [4,8,9]. It can learn from and adapt to new information without being explicitly programmed [10]. The ability of these algorithms to detect meaningful patterns has led to their adoption across various applications in sciences and technology, spanning from organic synthesis [11,12], and material science [1,13-15] to fabrication process optimization [16,17].

However, the discovery of new materials with optimized properties for semiconducting device functionality [17-19] has not yet been achieved. Particularly in the field of emerging photovoltaics, it has not been possible yet to invert the relation between device performance and a material´s structure due to the complex correlation between the material's structural features, the processing-dominated microstructure of composites and the relative impact of both on device performance. Prior efforts have focused primarily on using ML to optimize the fabrication process or to predict device performance and stability based on fabrication processes. The preliminary exploration has already been conducted in our laboratory, where Gaussian process (GP) regression was employed to model data from robotic device fabrication, enabling the analysis and prediction of device performance and stability [20-23]. The best parameter set and objective function could be quickly identified over the entire parameter space with a minimal number of samples. A similar approach was applied by Xu et al. to optimize the passivation materials for perovskite [24]. However, the training data is limited to the fabrication process or commercial materials, and as a result, it does not include the generation of new molecular structures. Assisting scientists in predicting and discovering new high-performance materials optimized according to the functional requirements of complex semiconductor devices is still a crucial challenge for ML-based materials discovery. Two recent studies have advanced the field by combining ML and organic synthesis. Bai et al. reported a successful case in which a gradient boosted tree regressor model, trained on data from 170 synthesized conjugated polymers, accurately predicted high-performance photocatalysts from a virtual database of 6354 candidates [25]. The trend obtained from the training data set was verified by experimental data on the newly synthesized polymers, leading to further refinement of the model. However, the insoluble characteristics of those polymers, while reducing the challenges of purification and enriching the database, have limited the broader applications of this material class. Gómez-Bombarelli et al. integrated virtual screening, cheminformatics, and organic synthesis to predict emitters for organic-light-emitting diodes [18]. At that time, the overall data volume and iteration number within the closed-loop were restricted by the number of synthesized molecules. Few general findings have emerged from these most recent studies, namely that the autonomous optimization algorithms require sufficiently large data volume but also data diversity, which necessitates the possibility of synthesizing structurally diverse molecules. Given the multidimensional nature of optimizing a chemical structure for device performance, the biggest challenge today is the generation of sufficiently large and consistent data sets to ensure the implementation of these algorithms [1,26,27].

To tackle these problems, we have developed a high-throughput (HT) organic synthesis platform, which can synthesize and purify more than 100 solution-processable small molecule semiconductors with varying structures and consistent quality over multiple synthesis campaigns within the shortest time [28]. This provided us with sufficient high-qualitative data for training an ML model coupling the structural features of the HTM to the performance of corresponding p-i-n PSCs. We sought to evaluate whether ML can be applied to the scale of data available to modern HT experiments and enable HTM discovery in multidimensional chemical space. To this end, we propose a workflow that couples automatic HT experiments, ML and validation through further HT experiments. These feedback loops will recommend new structures based on device results and allow for a closed optimization of the material to the target criteria of the solar cell. Due to the means of rapidly acquiring a substantial quantity of molecules, multiple iterations of experimental data based on organic conjugated molecules have been successfully carried out for the first time. The model, trained by 149 synthesized molecules, accurately predicted high-performance HTMs from a virtual database of $10^6$ candidates. Due to the already considerable complexity of this study, we did not undertake an individual optimization of the device structure for each new HTM during the BO operation. Instead, device optimization was done for a series of HTMs that showed an initial power conversion efficiency (PCE) exceeding 20% after BO operation. These materials finally reached efficiencies of over 26%, highlighting the enormous potential of coupling material and device optimization in a combined workflow. We believe that the relevance of this work goes far beyond the discovery of optimized HTM for PSCs.

**Workflow**
In this study, Suzuki coupling was used to synthesize new HTM molecules for perovskite solar cells. This reaction facilitates the combination of two distinct monomers, an A and a B type molecule, into a B-A-B type conjugated molecule. The workflow (Fig. 1) began with (i) the creation of a source database and the definition of sub-databases. The source database combines all commercially available monomers A and B compatible with Suzuki coupling bromines with boronic acids. The intermediate database, containing DFT calculations, consists of 13,000 randomly selected molecules from the source database. The synthetic database is then selected from the intermediate database according to specific rules (Kennard-Stone for the initial database and Bayesian for iteration database). The initial pre-training was executed for representative monomers (33*13) that were selected to ensure the broadest coverage of chemical features; (ii) DFT calculation of robust chemical descriptors were performed for each molecule; (iii) followed by high-throughput organic synthesis in 4 campaigns, purification and characterization, initially creating a database of 101 molecules (iv) processing & characterizing p-i-n perovskite devices from the 101 molecules (v) the application of machine learning based on calculated descriptors and experimental data for predicting the performance of each new molecule within the virtual compound library. (vi) subsequently, the model was validated by synthesizing 48 molecules suggested by the BO in two further campaigns from a space covering 50 A and 19 B molecules. This process can be iteratively repeated to explore further molecules. Finally, the set of structure-property relationships connecting device performance to molecular properties was built and analyzed by Multi-Task Gaussian process regressors (MTGPR), combining data from HTM films as well as HTM/perovskite interface characterization. Based on those rules, molecular design principles were established, leading to the iterative refinement of the molecular structure.

**Initial library generation -- synthesis and device manufacturing**
As the first step, the formulation of an in-silico library encompassing $10^6$ HTMs utilizing the RDKit package (see section S3 supplementary materials for details) is described. To amass a substantial dataset, we set two primary criteria for monomer selection: 1. reacting units (Br for monomer A, boronic acid derivatives for monomer B) attached to aromatic moieties; 2. monomer B is restricted to possessing only one reacting unit to prevent polymerization. Apart from this, no other selection criteria or "intuitive" selection rules were applied. Following these criteria, 1132 monomers A and 850 monomers B were selected. Subsequently, a script based on RDKit was employed to annotate the monomers across seven aspects, encompassing types of conjugated frameworks, substituent types, electronic effects, and steric effects (Fig. 2A). The categorization of HTMs along these seven aspects is aligned to the current state of understanding how HTMs impact perovskite device performance [7,29-31]. A representative library of monomers (33 A and 13 B) was then chosen from this space using the Kennard-Stone algorithm (Fig. 2B, 2C and 2D) [32]. This sampling method choice ensures that monomers are selected from uniform regions of the feature space. To guarantee comparability with the reported molecules, a few monomers with good reported performance were manually added. To make more efficient use of the selected monomers and compare the differences in their structure-performance correlation, we refrained from simple pairings. Instead, we divided the monomers into two groups for synthesis: A1-A30 combined with B1-B2, and A30-A33 with B2-B13. A comprehensive description of the synthesis, purification, and pre-characterization was recently published [28]. That routine is an essential pillar for the next step – the autonomous material discovery for optimized device performance – which we report in this work. In essence, we adopted a semi-automated synthesis platform where a microwave reactor accelerated the synthetic process. Subsequently, the synthesized molecules underwent a two-step purification involving fast filtration and recrystallization. Our HT workflow features the generation of 24 molecules within a week, including a most complete data acquisition. To ensure immediate access to the identified promising candidates and to assess the potential impact of synthesis and purification methods on material performance, the batch-to-batch reproducibility of the platform was verified by characterizing 10 representative molecules from different batches, including nuclear magnetic resonance (NMR), mass spectrometry, optoelectronic properties, and device performance (see section S2 in the supplementary materials). Minor fluctuations were observed between the different batches, guaranteeing that the reproducibility of our experimental data set does not limit the functionality of the ML algorithms. Such high reproducibility is further relevant for FAIR-guided data libraries, allowing other scientists to reproduce our experiments.

These synthesized molecules were employed as dopant-free HTMs in p-i-n structured PSC devices, where the function of the HTM goes beyond merely extracting and transporting holes; it plays a crucial role in influencing the crystal growth of the perovskite [29,33,34]. Here, perovskite is used as an absorber layer, ([6,6]-phenyl-$C_{61}$-butyric acid methyl ester (PCBM)/ bathocuproine as electron transporting layer and Ag as the top electrode. The resulting current density-voltage (*JV*) curve is depicted in fig. S9. Reference devices based on Poly(triarylamine) (PTAA), a state-of-the-art commercial HTM for p-i-n PSCs, were employed to calibrate the variations within each batch of devices. Furthermore, 6-12 devices were prepared per molecule to provide sufficient statistics on the data relevance. The average value was used as the experimental data point. To further ensure the validity of the data, reference devices were extensively optimized to minimize inter-batch differences. The device performance parameters of each novel molecule were normalized to the reference device with a PTAA hole transport layer before entering them into the ML model. Additionally, throughout the iterative process, the normalization to PTAA allowed us to adapt the perovskite recipe throughout the experimental campaign to achieve better reproducibility at the highest performance.

The heatmap of the PCEs for all synthesized molecules in Fig. 2E uncovers the first interesting trends. The PCEs of molecules AxB1 are generally higher than AxB2. Similarly, while keeping the monomer B constant, almost all molecules based on A30 as a central building block demonstrate the best performance. This signifies the substantial advantage of B1 and A30, triphenylamine (TPA) derivatives, in HTMs. However, there are notable exceptions: 1. A30B2 > A30B1, 2. A31B12 > A30B12, 3. A31B13 > A30B13. The first and second points share a commonality - the monomers B1 and B12 have a TPA structure. This suggests that excess TPA or its placement on the molecule's periphery may be disadvantageous. As for the third point, it appears challenging to explain it from chemical intuition alone. Utilizing machine learning is necessary to provide further insights and identify underlying mechanisms.

**Machine learning models and feature engineering**

To better understand structure-property relations in the observed data, we constructed an ML model that correlates representative molecular descriptors to the PCEs of the devices. In contrast to categorical labels such as A and B, continuous molecular descriptors can be used to provide an ML readable description that can integrate unseen A/B fragments into the same ontology. Typically, this process requires feature engineering to find meaningful descriptors for the problem at hand. To this end, a large virtual molecular library was generated by evaluating the expected reaction products from a set of A and B building blocks with the RDKit software package. Subsequently, the 3D geometry of the molecules was optimized using a conformer search in CREST and the semiempirical density functional tight-binding program xTB [25,35], followed by the calculation of molecular descriptors using DFT in TURBOMOLE (see section S3 in the supplementary materials for details). For the ML model, we sought a set of descriptors that adequately capture device differences without relying on a specific hypothesis. To achieve this, we chose a combination of simple molecular statistics, such as the number of atom species, aromatic bonds, and specific functional groups, with theoretically computed features such as the log solubility, molecular orbital energies, dipole moment and geometric properties such as rotational constants. A list of all features with explanations can be found in the supplementary materials. In Fig. 3A, we show the distribution of the most important features, such as the DFT highest occupied molecular orbital (HOMO) level, which should relate to the device performance in a step-function type from basic physical considerations. For model selection, we trained different ML models on a random 10-fold cross-validation of the 101 experimental molecular data points. Tested ML models include random forest regression, linear regression, neural networks, Gaussian process regression (GP), and kernel-ridge regression (Fig. 3B). All simple models perform equally well. For Bayesian optimization we chose the Gaussian processes as a surrogate model since it offers an uncertainty measure required in many acquisition strategies. The prediction accuracy of all device labels beyond PCE for the GP model is shown in Fig. 3C. A more detailed analysis of the influence of features and what physical insights can be learned from the ML models is discussed in the section *model analysis*.

**Experimental validation of the model**

To demonstrate that the ML model can indeed discover new molecules by predicting viable new organic semiconductors for hole extraction, we conducted two iterations of closed-loop materials optimization. This entails the identification of potential candidates via the ML surrogate model and Bayesian selection criteria, automatic synthesis of candidates, and finally, device characterization used to update the model. In the first iteration, 24 new molecules (blue triangles in Fig. 4A) were synthesized in a single batch and characterized to validate the previously obtained model. New building blocks (A525-pyridinone, A772-dicyanovinyl) and asymmetric structural motifs were considered to enrich the diversity of molecular structures in the database. The monomer database for synthesis was expanded from 33*13 to 50*19, mainly through machine learning recommendations, supplemented by accessibility, derivatives exhibiting high-performance structures in the initial database, and random acquisition.

The target for optimization was the device PCE averaged over 12 devices with an upper confidence bound (UCB) acquisition function, which is supposed to identify high-performance HTMs for further exploration. We found that the new series of materials resulted in device efficiencies that were generally higher than those of the materials in the initial database, demonstrating the significant advantage of machine learning over random sampling or grid search approaches when operated in the "exploit" mode. Among them, 6 molecules exceeded the PTAA-based device reference. We also observed outliers to the predictions (samples 110 and 120), which were separately examined (see section S4 in the supplementary materials) and could be explained in terms of impurities, unfavorable wetting behavior or limited solubility.

For the second iteration, we increased the explorative character of the UCB in the model [12,23]. Subsequently, molecules 126-149, recommended by ML, were synthesized (Fig. 4). Although no new champion HTM was found, the average PCE of predictions was still very high, comparable to the first iteration. This is a very strong affirmation of the potential and viability of our workflow, considering the multitude of factors governing the performance of perovskite devices. Device performance is only partially limited by the HTM, as many other layers, including the electron transporting layer or the electrodes as well as the quality and defect density of the absorbing layer, are known to reduce performance. Moreover, the physical models behind these losses cover multiple chemical and physical phenomena, including (i) inherently limiting material properties, (ii) generation, recombination and transport dynamics in semiconductors, (iii) the energetics of interface formation and the corresponding potential landscape in thin film devices, (iv) thin film microstructure formation (v) the device architecture itself and (vi) the macroscopic film homogeneity. All of that is described in the ontology of perovskite devices which has been recently uploaded on Matportal [36]. Under these aspects, we were surprised to see the spread in efficiency among the various HTMs within our library.

Clearly, we expected a threshold type behavior for HTM materials having a too small bandgap or a too small or even negative offset in the HOMO / LUMO potentials vs. the valence band maximum (VBM) and conduction band minimum (CBM) of the perovskite. However, the broad variation of representative performance values ranging from 15% to 21% postulates that the nature of the HTM molecules is capable of determining the performance of perovskite devices beyond our expectations. It's worth noting that to obtain accurate trends, the performance of these molecules was characterized under uniform device conditions, with the standard device optimized based on PTAA. When exploring the limit of a material's performance, device parameters need to be adjusted according to the material's characteristics. Based on this, three promising molecules were selected for individualized device performance optimization. We adopted a bottom-up optimization approach, from HTM to perovskite to electron transporting material (ETM) (see section S5 in the supplementary materials). Under these efforts, the device efficiency improved by 10% - 20%, reaching performance levels of 23.5% - 24.3% with FF of 87%. Considering that new conditions might affect the original ranking of materials, thereby influencing ML predictions, we selected 20 molecules from the database to observe the impact of individualized optimization on trends. The results showed that it slightly affected the open-circuit voltage ($V_{oc}$) and fill factor (FF) trends, but the impact on the PCE trend was negligible (fig. S30). To further fully explore the potential of those materials on $V_{oc}$ and $J_{sc}$, we further refine the perovskite formulation and ETM to better align with the properties of those HTMs (Fig. 4E). As a result, we observed a significant improvement in the performance, achieving a $V_{oc}$ of 1.195 V, a $J_{sc}$ of 26.01 mA/cm$^2$, a FF of 84%. These enhancements culminated a PCE of 26.23% (certified 25.88%) alongside an excellent operational stability, maintain 80% of the initial performance ($T_{80}$) for over 1000 h. Finding performance values as high as 26.23% for a process, in which the state-of-the-art material MeO-2PACz is reaching an efficiency of about 24% (Supplementary Fig. 32-34), demonstrates the power of this approach. Moreover, it outlines the next major steps for the combined material and device acceleration strategy – that has to be walking away from single-objective optimization tasks to multi-objective optimization tasks. In terms of our work, we see the next major challenge as the simultaneous optimization of the efficiency and operating life of perovskite solar cells as a function of the HTM.

**Model analysis**
To obtain interpretable insight into what our ML models have learned and to identify physical parameters that influence the device performance, we added further experimental material properties and extracted feature importance information from the trained ML models (Fig. 5). Since the HOMO position is relevant for charge extraction from physical considerations, the HOMO level position is expected to be aligned between the electrode and the perovskite, and is identified as an important feature in Fig. 5A [37-39]. In order to identify more decisive features, we conducted a feature analysis using the Recursive Feature Machine (RFM), a kernel machine that recursively learns features importance [40]. As in Radhakrishnan *et al.*, we used a generalization of the Laplacian kernel that incorporates a learnable feature matrix $M$ to compute the Mahalanobis distance between data points [41]. The $R^2$ value on the test data of this method was evaluated to be approximately 0.5. On 100 randomized train-test splits, we ranked the largest feature-matrix values of the trained Kernel methods in Fig. 5C. The features, the RFM model focuses on, are purity, HOMO level, HOMO/LUMO gap, and the presence of tertiary amines. Besides the electronic properties of the molecule, the purity of the synthesis product is the most crucial descriptor for the final device performance. This corroborates that impurities typically reduce overall performance due to potential diffusion and the introduction of traps or unwanted doping in adjacent layers. Other important factors are the presence of nitrogen atoms, which reflects the observations mentioned above in Fig. 2F, dipole moment, molecular shape, and to a lesser degree composition and overall bond type or conjugation.

Additional experimental input from extended characterization was generated to increase the diversity of input/output parameters, including wettability, photoluminescence quantum yield (PLQY), and time-resolved photoluminescence (TRPL) (see section S1 and S4 supplementary materials for details). Since these measurements are often only available after producing the device, we added them as ulterior outputs and trained a multi-task GP model (MTGPR) on the joined target space. The task covariance matrix represents potential correlations that the MTGPR model uses with a kernel function shared among tasks (Fig. 5C). It shows the expected correlation between $V_{oc}$, $J_{sc}$, and FF with the PCE. Furthermore, to a lesser extent, it also indicates correlations between PCE and additional labels, e.g. with TRPL, which would be consistent with current reports [42-45]. However, the correlation is not as pronounced as expected and does not lead to a statistically significant improvement in PCE prediction accuracy when the additional labels are fed to the multi-task model.

Since the dataset is not strictly i.i.d., we additionally evaluated the generalizability of our ML models in a leave-one-out cross-validation scheme. We iteratively picked each single molecule as a test set and removed the same A and B building blocks from the training set, in order to make sure that the model cannot produce its predictions by simply learning to recognize the molecular fragments. We trained GP and RF models and evaluated their test error with this validation scheme, reaching in both settings an $R^2$ value of approximately 0.3. Although the test error increased, the model could still generalize to unseen data points. The most relevant features in this setup remained similar to Fig. 5A in the importance ranking: The presence of tertiary amines, HOMO/LUMO gap, and HOMO level, but also dipole moment and the presence of biphenyl substructures. We can confirm that the model learned to predict the perovskite device performance for unseen HTM molecules, as required for materials discovery.

To have a more interpretable model, we trained a linear regression model applying first both a forward and a backward sequential feature selection, which is a family of greedy search algorithms used to reduce the feature space to a lower subset. We evaluated the resulting models with the Bayesian information criterion (BIC) to select the best-performing set of features we had learned so far. Our selected model uses eight features (aromatic bonds and atoms counts, logP, count of nitrogen atoms, purity, dipole, rotation constant c, and the presence of tertiary amines) to predict the PCE, achieving an $R^2$ of approximately 0.46 (Fig. 5D), which is higher than any other model we found.

Finally, in order to provide chemists and material scientists with a clearer understanding of our findings, enabling them to delve deeper into molecular design based on our finds, here, we employed chemical language to elucidate the results of ML (Fig. 5B). The feature importance plot distinctly highlights the significance of HOMO and tertiary amine in the model. The significance of HOMO in molecular design has been widely recognized due to its decisive role in charge extraction at interfaces. However, tertiary amine is often overlooked. Upon examining all synthesized molecules, we discovered that tertiary amine often refers to triphenylamine (TPA), whose low ionization potential contributes significantly to the molecule's HOMO [6]. Based on these two descriptors, all synthesized molecules can be categorized into three types: Type I, TPA-absent molecules, referred to as AxBy; Type II, with TPA on the periphery, typically AxB1 structures; and Type III, with TPA at the molecular center, typically A30By structures. Under this classification, a pattern emerges in HOMO and PCE: (I) HOMO ranges from 5.1-6.1 eV with 5%-14% PCE; (II) HOMO from 4.3-5.2 eV with 13-20% PCE; (III) HOMO from 4.9-5.7 eV with 15-21% PCE. This classification narrows the candidate pool from $9.6 \times 10^5$ to $5.8 \times 10^3$ molecules. Once the A position is established as TPA, molecular properties are primarily influenced by the B-position group. The feature importance analysis also highlights the roles of the HOMO/LUMO gap and dipole moment. The combination of TPA and acceptors ensures an appropriate HOMO-LUMO bandgap, with heteroatoms in acceptors also contributing to perovskite passivation. This combination further reduces candidates from $5.8 \times 10^3$ to $4.6 \times 10^2$. To facilitate rapid selection by chemists without DFT calculations, we used the Topological Polar Surface Area (TPSA) as a rough indicator of the building block's polarity and electron-withdrawing capacity, readily searchable via PubChem. Finally, the performance of molecules should be finely tuned based on TPA+acceptors, such as the edge-on orientation positively impacting passivation and charge transport. In the combination of TPA derivative and 5 B groups (fig. S22), device performance is systematically enhanced through fine-tuning of the B-position group and TPA structure. For instance, groups like A770, with slightly weaker symmetry, tend to exhibit better device performance. This fine-tuning can reduce the number of candidate molecules from $10^2$ to $10^1$, a quantity that is well within the realm of high-throughput synthesis. We summarize by highlighting the two-fold strategy learned from training an ML model to be capable of predicting such a complex property as a device performance based on molecular structure input. Such model can be further explored in a two-fold strategy. On the one hand, it can be used in autonomous workflows to identify and predict further novel molecules. On the other hand, synthetic researchers can use that model to predict perovskite device performance for new molecular designs within a certain chemical space. That can be further guided and supported by the set of design rules extracted that are elucidated from a fully trained model.

**Conclusion**

We believe this work transcends the current strategy of discovering new materials focused solely on one specific property. Including organic synthesis in self-driven autonomous labs in combination with autonomous device optimization lines is a huge step towards fully autonomous materials discovery in many application areas. Here, for the first time, we demonstrate a unique workflow that has the power to discover functional materials optimized for highly complex applications such as photovoltaic devices. A significant challenge we solved was to build predictive models based on molecular descriptors, allowing us to link the structure of a material to the performance of a highly complex device such as a solar cell. This is particularly important when it comes to device processing and optimization, which necessitates a nuanced understanding of both the material and the process involved. This aspect can constitute a breakthrough for overcoming the structure–process–property nexus for disordered materials.

Looking forward, we aim to integrate material discovery and device optimization into a seamless, closed-loop process. Achieving this will require a concerted effort in interdisciplinary research, combining insights from material science, engineering, and advanced computational techniques to create a synergistic workflow. This integrated approach is the most promising strategy to revolutionize the way we develop and optimize materials for cutting-edge technological applications.

**Legends of the main text figures**

**Fig. 1. Approach overview. I. Database.** There are three kinds of databases. 1. Source database: the virtual combination of two types of commercial monomers using Suzuki coupling rule. 2. Intermediate database: Randomly selected molecules from the source database for DFT calculations. 3. Synthesized database: Synthesized molecular collections in this study, including an initial database for model training and two iteration databases for model validation and correction. **II. DFT calculation**. Descriptors of molecules in the intermediate database are calculated via DFT. **III. HT Synthesis and characterization**. Molecules in the synthesized database are synthesized, purified, and characterized through our in-house high-throughput platform. **IV. Device and semi-device data**. The synthesized molecules are used as HTM in PSCs and characterized in devices and semi-devices (HTM/active layer). **V. Model training and iteration**. The model is trained based on HTM descriptors and device parameters. The new promising molecules are predicted, synthesized and experimentally measured and fed back to the database. The iteration is repeated until the discovery of the best HTM. **Molecular iteration**. Material design principles are summarized and analyzed.

**Fig. 2. Generation of the initial library.** (**A**) Monomer descriptor calculation from different perspectives. I. Aromatic ring species; II. Conjugate length; III. Substituent species; IV. Active group; V. flexible and rigid units; VI. Electronic effect; VII. Spatial effect. (**B**) Monomer subset selection (blue ball) with the Kennard-Stone algorithm from a commercial monomer library (square). (**C**) and (**D**) Selected monomers for the initial database. (**E**) Color map of PCEs for HTMs in the initial library. It consists of two parts: the combination of A1-A30 with B1-B2 (upper part) and A30-A33 with B2-B13 (lower part). In the upper part, the serial number of each monomer A is filled in the cell. (**F**) In-silico library for HTM descriptors used for initial library selection.

**Fig. 3. Model training based on experimental data and in silico descriptors.** (**A**) Distribution of important features from the dataset used in the model training. (**B**) Comparison of the validation error of different models to compare their performance. (**C**) The prediction accuracy of the Gaussian Process (GP) for all device labels, including maximum PCE, $J_{sc}$, $V_{oc}$ and FF.

**Fig. 4. New synthesized molecules and experimental data for iteration.** (**A**) Predicted properties of the molecular library. (**B**) Selected molecules from the database for iterations based on calculated properties, where marker size is proportional to device performance. Blue triangles represent the molecules in the first iteration and the orange ones in the second. (**C**) Molecular fragments used of iterative molecules. To clearly visualize the structure of the final molecules, molecular fragments are used to represent the monomers. (**D**) Device performance of molecules for the initial dataset and the iterations. The PCE of the samples is the average performance normalized to the reference devices (based on PTAA) of the corresponding batch. (**E**) Further personalized optimization of the devices based on representative molecules. Blue: standardized condition. Red: personalized condition based on molecular properties.

**Fig. 5. Model analysis.** (**A**) Feature importance of the RFM evaluated from the M matrix coefficients. (**B**) Molecular design rules guided by machine learning results. (**C**) Covariance matrix of the multi-

label GPR for PCE and additional device characteristics. $t1_{perov}$ and $t2_{perov}$ represent the carrier lifetimes of two distinct recombination processes within the perovskite layer when the excitation laser is incident from the side of the perovskite. $t1_{glass}$ and $t2_{glass}$ correspond to data that laser incident from the glass side. $DV_0$ and $DJ_0$ are the derivatives of the *J-V* curve in $V=0$ and $J=0$. **(D)** Linear model test predictions for the leave-one-out CV scheme.

# Methods

## Materials

Reagents and solvents for organic synthesis were purchased from commercial suppliers (Fluorochem, Sigma-Aldrich, BLD pharm, TCI Europe, Apollo, Alfa Aesar) and used with no further purification unless otherwise noted. Thin layer chromatography (TLC) plates were purchased from Sigma-Aldrich. Chemicals for perovskite solar cells: Formamidinium iodide (FAI), methylamonium bromide (MABr) and methylamonium chloride (MACl) were purchased from Greatcell Solar Materials. lead iodide ($PbI_2$), Lead bromide ($PbBr_2$) and Lead chloride ($PbCl_2$) were purchased from TCI chemicals. Cesium iodide (CsI), *N,N*-Dimethylformamide (DMF, anhydrous), isopropyl alcohol, dimethyl sulfoxide (DMSO, anhydrous), and chlorobenzene were purchased from Sigma-Aldrich.

## High through-put synthesis

The amount of theoretical products is 0.2 mmol for most reactions, except for some products with too large or too small molecular weight. We set the weights of theoretical products around 100 mg. The stock solution-1 dissolving $Pd(OAc)_2$ (0.01 mmol) and SPhos (0.02 mmol) in dioxane, stock solution-2 of 3 M $K_3PO_4$ in $H_2O$, were firstly prepared and degassed with $N_2$. Then monomer A (0.2 mmol, 1.0 equiv.) and monomer B (0.8 mmol, 4.0 equiv.) were added into vials and sealed under the inert atmosphere. 3 ml stock solution-1 and 0.5 ml stock solution-2 were injected into vials with a syringe through the septum. Before the start of the heating program in the microwave reactor, we set a pre-stirring for 5 min to dissolve the starting materials. Then the reaction temperature was raised to 90 °C and kept for 30 min. Here, we set the power of the microwave limit not to exceed 100 W. The reaction mixtures were cooled to room temperature. Aqueous layers were removed manually with pipettes. The remaining organic layers were transferred into the SEP tube of the vacuum manifold. After filtering, the SEP was rinsed 5 times with 2 ml THF. The solvents were removed by a sample concentrator, and the residues were purified by recrystallization with optimized mixed solvent and recrystallization again if needed by encoding.

## Device fabrication

Perovskite solar cell devices having the structure of ITO/HTM/perovskite/PCBM/BCP/Ag were fabricated, where the Indium tin oxide (ITO) was the bottom layer. The patterned indium tin oxide (ITO) glass substrates are cleaned for 15 minutes with deionized water, Acetone, and Isopropanol in an ultrasonic bath. The layers are then dried well before treated in an UV-ozone cleaner for 15 min. After UV-ozone treatment, the ITO substrates immediately are transferred to $N_2$ filled glovebox. HTM (2 mg/ml in chlorobenzene) was spin-coated on the ITO substrate at 5000 rpm for 30s and then annealed at 100 °C for 10 min. Throughout the workflow iterations, we continuously updated the perovskite recipe. The article involves two specific recipes. Recipe 1 is 1.5 M $(CsI)_{0.05}(MAPbBr_3)_{0.17}(FAPbI_3)_{0.83}$, and recipe 2 is 1.45M (CsI)$_{0.17}$(FAPbI$_3$)$_{0.83}$. For recipe 1 (CsMAFA), PbI$_2$ (1.5M) and PbBr$_2$ (1.5M) are dissolved in a mixture of anhydrous Dimethylformamide (DMF): Dimethyl sulfoxide (DMSO) (4:1 volume ratio), respectively, and then added to formamidinium iodide (1.09:1 molar ratio) and methylammonium bromide (1.09:1 molar ratio) powders, respectively, to obtain FAPbI$_3$ and MAPbBr$_3$ solutions. These two solutions are then mixed in a 17:83 volume ratio. Finally, the cesium cation is added from a 1.5 M CsI solution in DMSO in a 5:95 volume ratio. For recipe 2 (CsFA), PbI$_2$ (668.4 mg), FAI (207.0 mg), CsI (64.0 mg), PbCl$_2$ (40.2 mg) and MACl (5 mg) are dissolved in 1ml DMF : DMSO mixture solution (4:1 volume ratio).

The perovskite solution is spin-coated on top of the HTM layer using the following program: 1000rpm for 10s, 5000 rpm for 30 s. After 25 s, 250 µl of Chlorobenzene is dropped on the spinning substrate. Then the perovskite film is annealed at 100°C for 10 min, and another higher temperature for 10min (120°C for recipe 1 and 150°C for recipe 2). PCBM solution (20 mg/ml in chlorobenzene) is spin-coated on top of perovskite at 1000rpm for 30s, then annealed at 80°C for 10 min. BCP solution (0.5 mg/ml in isopropanol) is spin-coated on top of PCBM at 5000rpm for 30s, annealed 80°C for 5 min. 100 nm thick Ag layer was thermally evaporated under a vacuum of $8\times10^{-6}$ mbar at a rate of ~0.1 nm/s to finish the device fabrication. The device active area was 0.08 cm$^{-2}$.

Individualized optimization of recipe 2 was conducted based on the final discovered promising molecules. The HTM concentration (0.5-5 mg/ml), antisolvent (chlorobenzene, toluene, and ethyl acetate), additives in perovskite precursor solution (PbCl$_2$, MABr), passivation materials and electron transporting layer were optimized (Figure S31-S38). The best condition is PbCl$_2$ 5 mg/ml + MACl 7.5 mg/ml + 145 °C annealing 10 min +N2200 doped PCBM.

Individualized optimization of recipe 3 (Fig. 4E) with a device structure of FTO/HTM/perovskite/C$_{60}$/BCP/Ag was conducted based on the final discovered promising molecules. 1.5 M Cs$_{0.05}$MA$_{0.1}$FA$_{0.85}$PbI$_3$ perovskite precursor was fully dissolved in mixed solvents of DMF and DMSO (4:1, v/v) with the molar ratio for FAI/MAI/CsI of 0.85:0.1:0.05. Then, 10 mg ml$^{-1}$ MACl was added to the solution to improve the film morphology. The precursor solution was filtered through a 0.22 µm polytetrafluoroethylene membrane before use. A portion of 70 µl of perovskite solution was deposited on the substrate and spun cast at 1,000 rpm for 10 s, followed by 5,000 rpm for 40 s. A 150 µl portion of chlorobenzene was dropped onto the substrate during the last 5 s of spinning, resulting in the formation of dark brown films that were then annealed on a hot plate at 100 °C for 20 min. For the devices with surface passivation treatment, a mixture of PEACl, 1 mg/ml)was dissolved in IPA dynamic spun on the as-prepared perovskite films at 5,000 rpm. for 30 s, then annealing at 100 °C for 10 min. The C$_{60}$ (30 nm) and BCP (7 nm) were deposited sequentially with a rate of 0.3 Å s$^{-1}$ and 0.5 Å s$^{-1}$, respectively, at a pressure of approximately $2 \times 10^{-6}$ mbar. Finally, Ag contact (100 nm) was deposited on top of BCP through a shadow mask with the desired aperture area.

**Film characterization**
A self-constructed high-throughput setup for time-resolved photoluminescence (TRPL) was used to collect the signal from the top side of the perovskite films with a 6.4µW laser power with an excitation wavelength $\lambda_{ex}$ = 402 nm. TRPL were fitted with a two-component exponential decay model. carrier lifetimes

$$I(t) = A_1\exp(-t/\tau_1) + A_2\exp(-t/\tau_2)$$

t1$_{perov}$ and t2$_{perov}$ in Fig.5 C represent the $\tau_1$ and $\tau_2$ in the formula, where the laser is incident from the side of perovskite. t1$_{glass}$ and t2$_{glass}$ correspond to data that laser incident from the glass side. Similar expressions also exist in PLQY data (PLQY$_{perov}$ and PLQY$_{glass}$). PLQY was calculated from the absolute PL measurements with integrating sphere, a 405 nm laser diode as the excitation source was adopted, and the laser intensity was altered by a step beam attenuator. The optical fiber was used to connect the sphere and a silicon CCD array detector; a 420 nm filter was used to increase the integrated time. *J–V* characteristics were measured with a Keithley source measurement unit and a Newport Sol3A solar simulator, which could provide illumination with an AM1.5G spectrum and light intensity of 100 mW cm$^{-2}$. The light intensity was calibrated with a standard crystalline Si device. The film's contact angles (CA) were measured on a Dataphysics OCA20 contact-angle system at room temperature. 5.0 μl water droplet was dropped onto the HTM films.

**Device characterization**

*J-V* curve of the solar cells was characterized using a Keithley source under 100 mW/cm2 AM1.5G illumination (Newport Solla simulator). The light intensity was calibrated with a crystalline Si-cell. The *J-V* data were collected from -0.2 to 1.2 V (forward scan) and 1.2 to -0.2 V (reverse scan) at a scan rate of 40 mV/s. An aperture mask with an area of 0.08 cm$^2$ was used.

**Data and materials availability**

All data are available in the main text or the supplementary materials. Code and data to reproduce the plots shown in the machine learning section of the manuscript are publicly available on GitHub (https://github.com/aimat-lab/perovskite_htm_screening).

**Acknowledgments** J.W. and J.L. acknowledge the financial support from the Sino-German Postdoc Scholarship Program (CSC-DAAD). C.J.B. gratefully acknowledges financial support through the "Aufbruch Bayern" initiative of the state of Bavaria (EnCN and "Solar Factory of the Future"), the Bavarian Initiative "Solar Technologies go Hybrid" (SolTech), and the German Research Foundation (DFG) SFB 953−No. 182849149. J.W., J.H., Y.Z. and C.J.B. gratefully acknowledge the grants "ELF-PV-Design and development of solution processed functional materials for the next generations of PV technologies" (No. 44-6521a/20/4) by the Bavarian State Government. J.Z., Z.X., and K.Z. acknowledge financial support from the China Scholarship Council (CSC). Y.Z. acknowledges the Alexander von Humboldt Foundation for supporting his scientific research during the postdoctoral period (Grant No. 1199604). Calculations were performed on the HoreKa supercomputer funded by the Ministry of Science, Research and the Arts Baden-Württemberg. P.R., L.T. and P.F. acknowledge support by the state of Baden-Württemberg through bwHPC. L.T. and P.F. acknowledge support by the Federal Ministry of Education and Research (BMBF) under Grant No. 01DM21002A (FLAIM). P.F. acknowledges support by the Federal Ministry of Education and Research (BMBF) under Grant No. 01DM21001B (German-Canadian Materials Acceleration Center). J.W. and J.S.R.O gratefully acknowledges the financial support from Solar TAP initiative by the Helmholtz Foundation/BMBF. J.L. gratefully acknowledges financial support from the National Natural Science Foundation of China (Grant No. 62104031) and the Technical Field Funds of 173 Project (Grant No. 24JJ210663A). M.H., B-w.P., and S.I.S. acknowledge financial support from the Basic Science Research Program (NRF-2018R1A3B105282023) through the National Research Foundation of Korea (NRF) funded by the Ministry of Science, ICT and Future Planning (MSIP).

**Author Contributions** J.W., J.H., P.F., and C.J.B. conceived the idea and supervised the project. J.W. performed the high-throughput synthesis and perovskite device fabrication. L.T., P. R., and P.F. applied the machine learning algorithms. M.H. and J.W. fabricated the perovskite device and wrote the manuscript. J.Z., K.Z., and Y.W. characterized the PLQY and TRPL. Z.X. developed algorithms to process UV-Vis data. J.S.R.O. provided partial materials. L.W., B-w.P., A.B., Y.Z., J.L., and L.L. analyzed the data. All authors reviewed and edited the manuscript.

**Competing interests** The authors declare no competing interests.

**Additional information**

**Supplementary Information** is available for this paper.
Correspondence and requests for materials should be addressed to Jianchang Wu, Pascal Friederich, or Christoph J. Brabec.

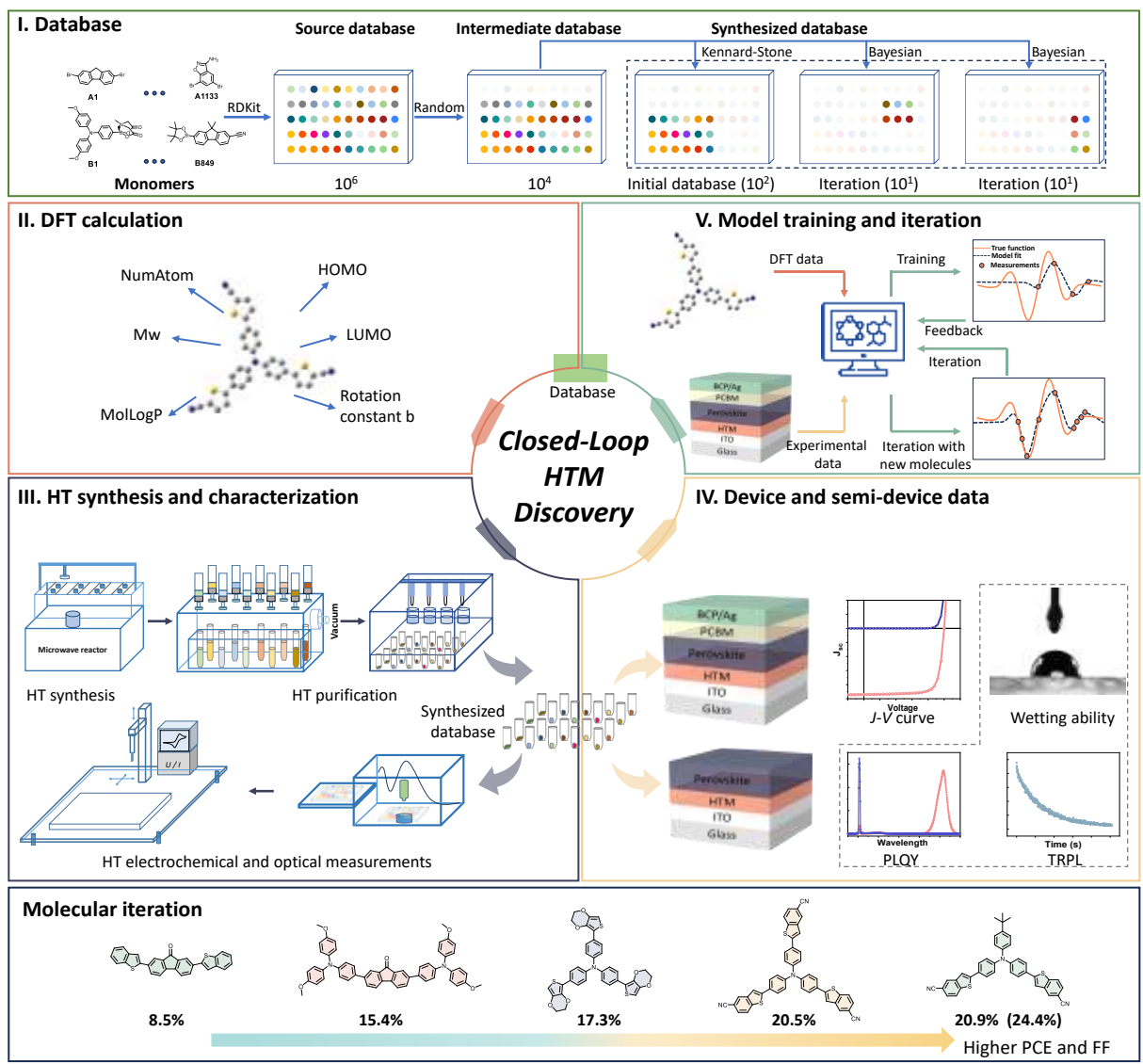

FIGURE 1

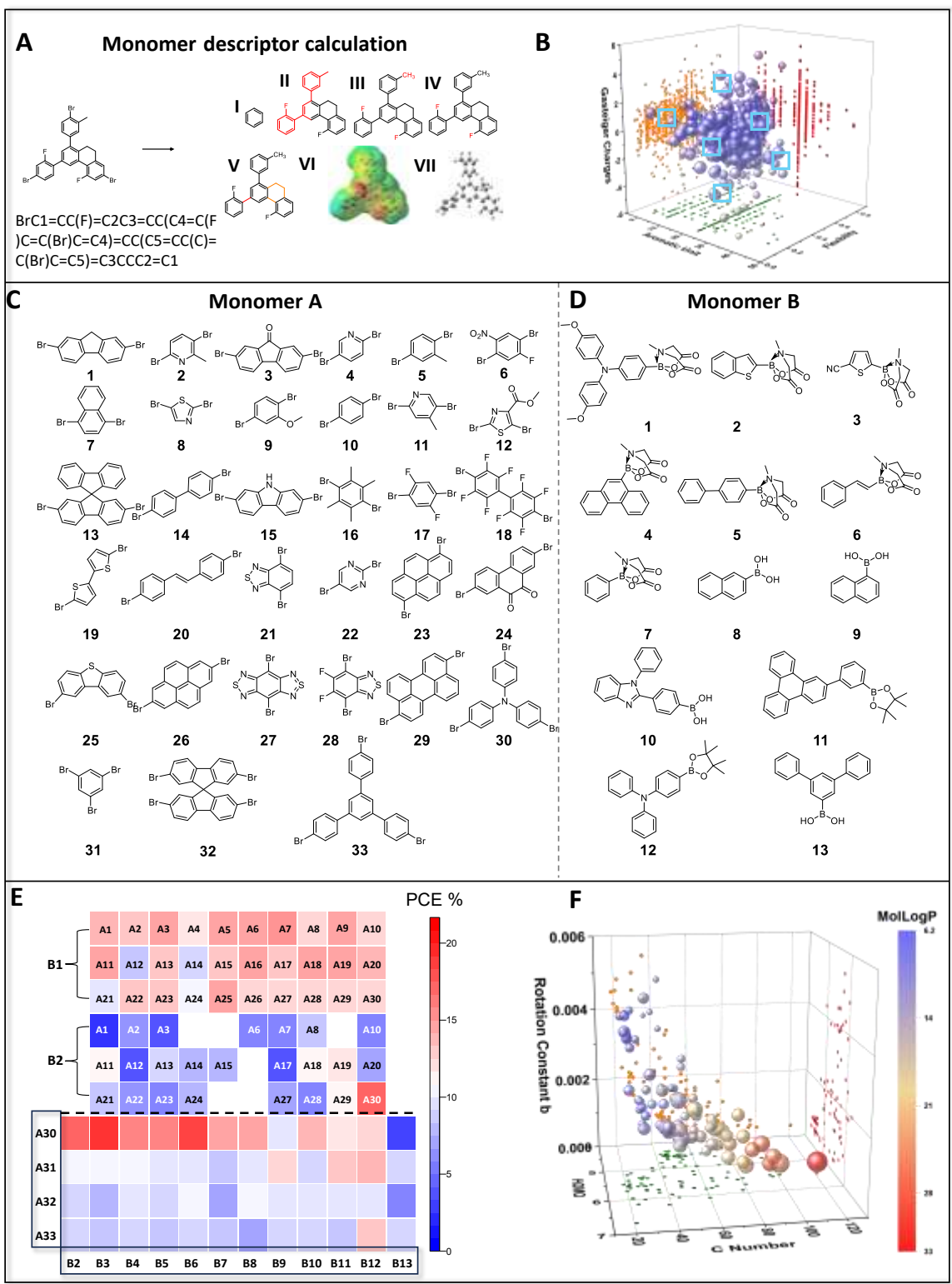

FIGURE 2

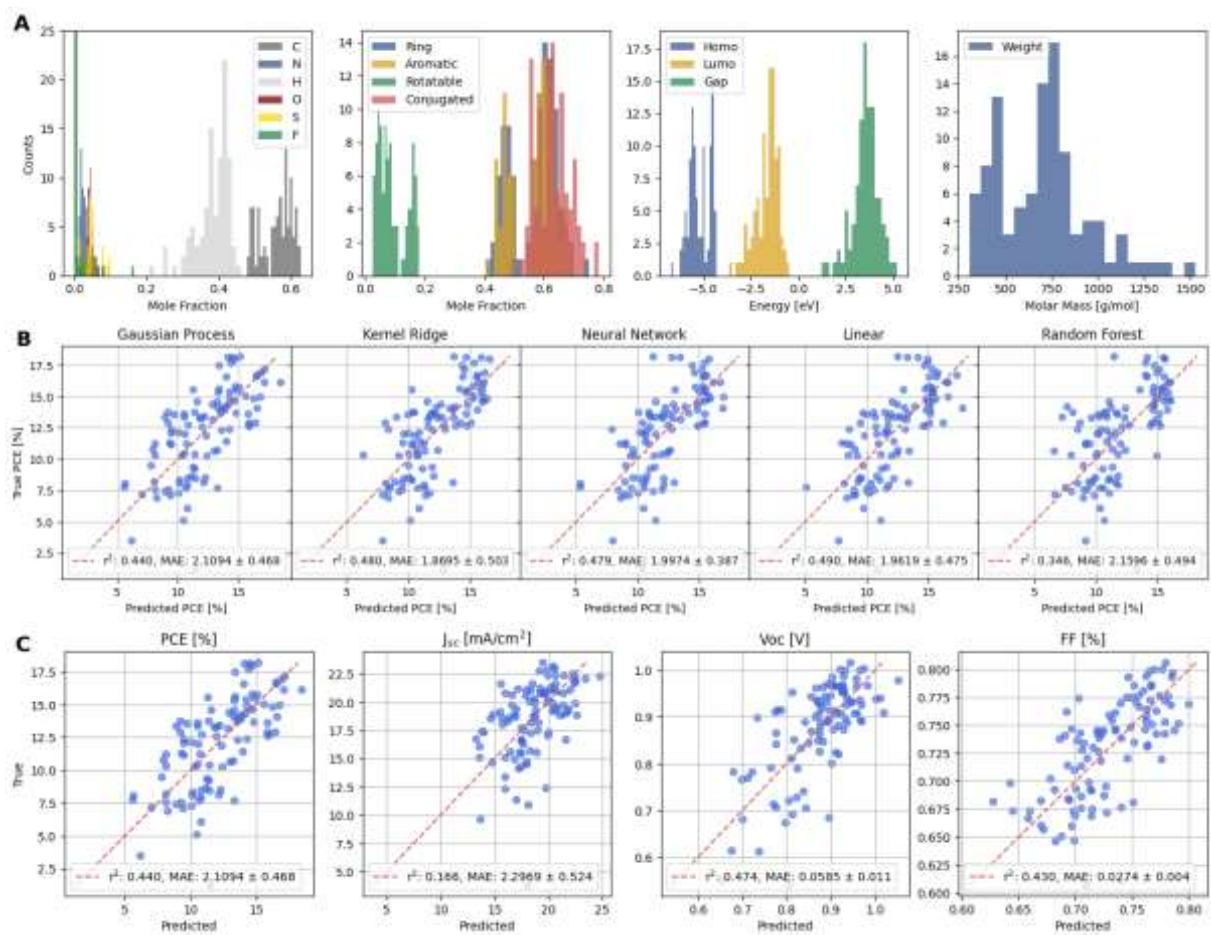

FIGURE 3

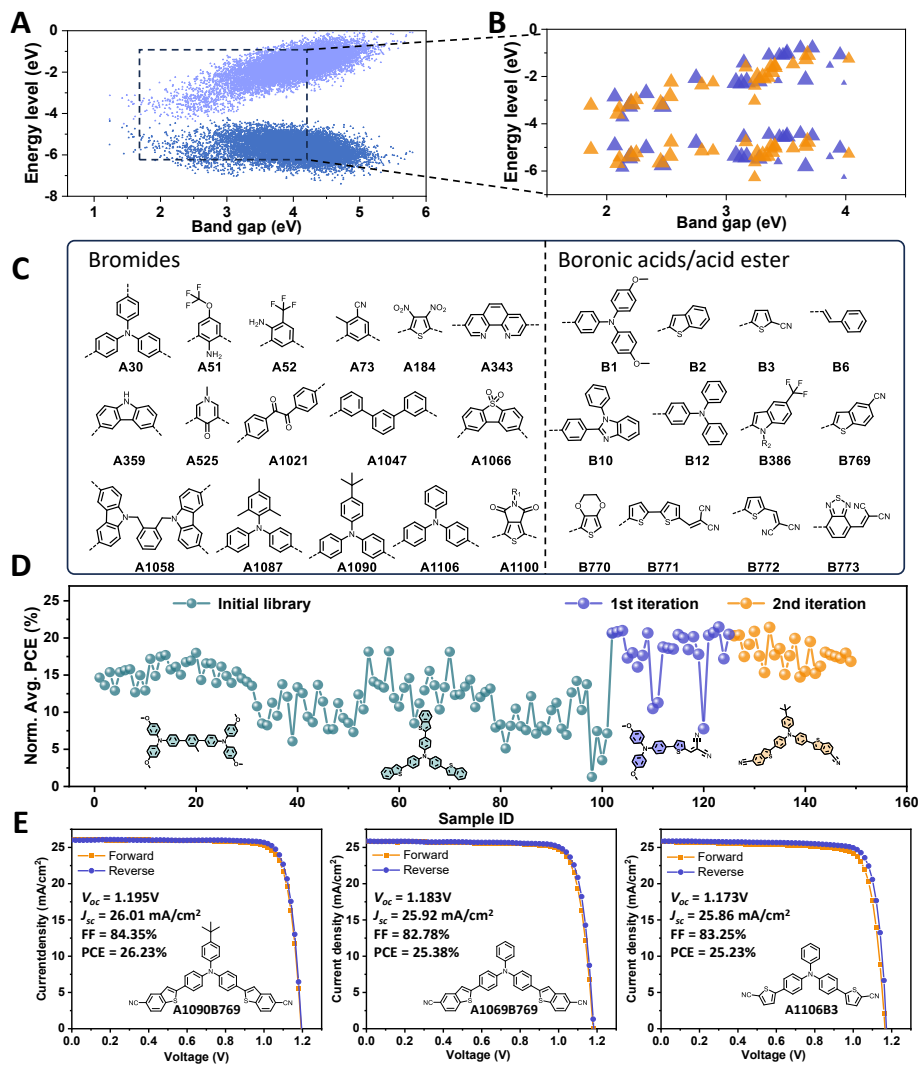

FIGURE 4

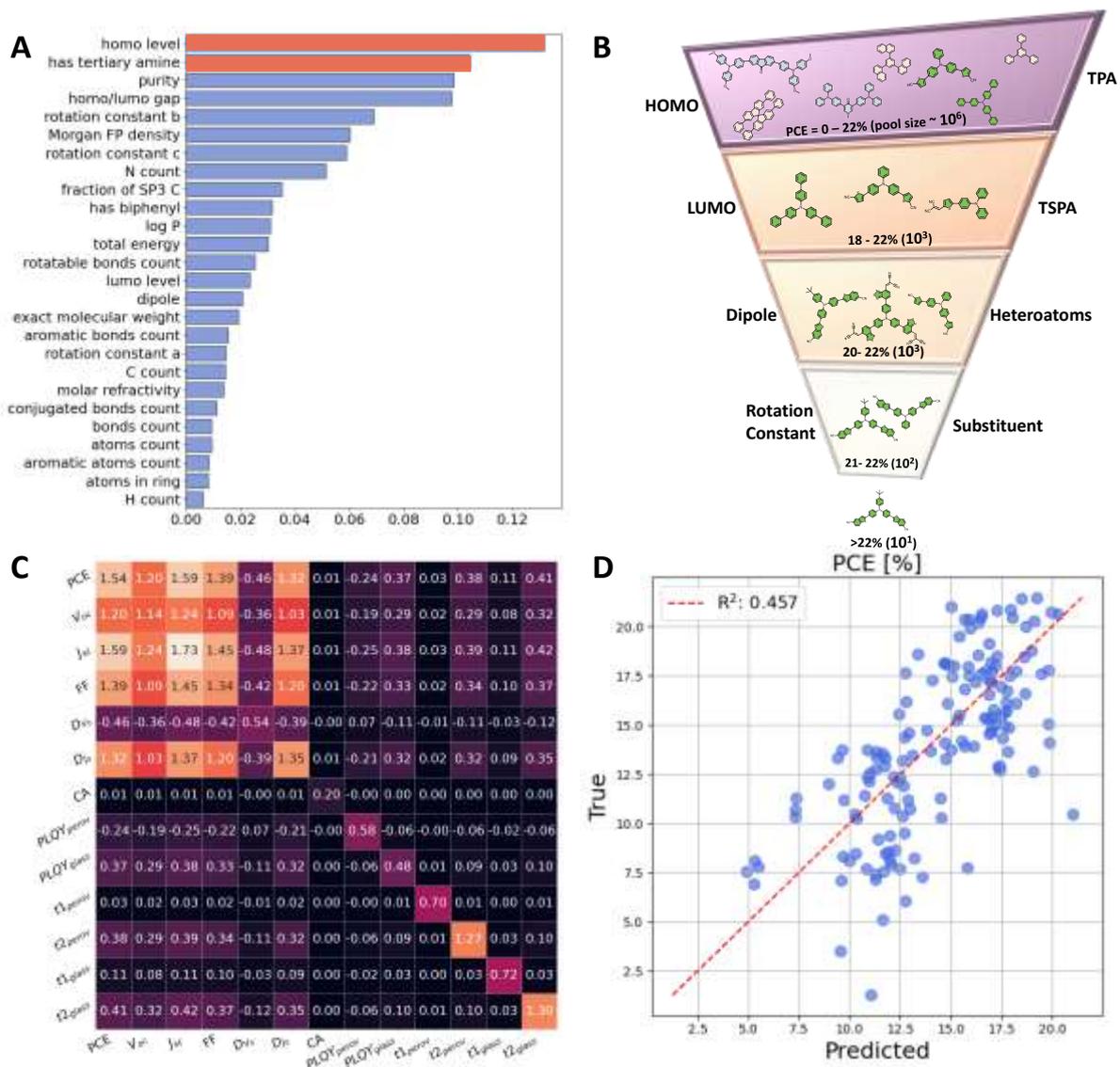

FIGURE 5